# Human being is a living random number generator

**Arindam Mitra**

Anushakti Abasan, Uttar Phalguni -7, 1/AF, Salt Lake,
Kolkata, West Bengal, 700064, India.

**Abstract**: General wisdom is, mathematical operation is needed to generate number by numbers. It is pointed out that without any mathematical operation true random numbers can be generated by numbers through algorithmic process. It implies that human brain itself is a living true random number generator. Human brain can meet the enormous human demand of true random numbers.

Number as well as a string of non/pseudo-random numbers can be generated by numbers with the help of mathematical operation through algorithmic process [1,2]. Although there is no mathematical proof that random numbers cannot be generated by any classical algorithm, but it is widely believed that random numbers can be generated only by using random events. It may be recalled that von Neumann first ruled out [1] the possibility of generating random numbers by any mathematical operation. There is a caveat. von Neumann did not rule out the generation of random numbers by any algorithmic technique. In fact, it is possible to generate random numbers by using simple quantum algorithm [3] which exploits intrinsic randomness of quantum system.

From the above discussion it may be conjectured that if any classical algorithm does not require mathematical operation to generate numbers, then it can be a good candidate for the generation of both random and non-random numbers. We have found such algorithm. Let us first present the most simple one.

## Algorithm- I

a). Take a string $S^m$ of 2n random bits.

b).Note down the random positions of 1s in the ascending order. It gives a sequence $K^r$ of positive integers. Similarly, note down the random positions of 0s in the ascending order. It gives another sequence $K^P$ of positive integers. The operation can be symbolically described as, $S^m \neg K^r$, $S^m \neg K^P$.

c). Take another sequence $S^f$ of 2n random bits.

d). Apply $K^r$ and $K^P$ on $S^f$, and thereby extract a pair of sequences $k^r$ and $k^p$ of random bits. It means bits of $k^r$ and $k^p$ are extracted from some random positions in $S^f$ determined by $K^r$ and $K^P$ respectively. That is, bits of $k^r$ are extracted from those positions in $S^f$ which are the positions of 1s in $S^m$ and bits of $k^p$ are extracted from those bit positions in $S^f$ which are the positions of 0s in $S^m$. The operation can be symbolically described as $S^m \neg K^r : S^f :: k^r$ and $S^m \neg K^r : S^f :: k^p$.

In this operation, two bits bb are used to generate a single bit b where the first bit belongs to $S^m$ and the second bit belongs to $S^f$. For $k^r$, $10 \equiv 0$, $11 \equiv 1$. For $k^p$, $01 \equiv 1$ , $00 \equiv 0$.

It is trivial to see that the algorithm generates a bit 0 or 1 with probability 1/2.

***Proof:*** Since $S^m$ is a sequence of random bits, *a priori* probability of getting 0 or 1 in any position is $p^m = 1/2$. Since $S^f$ is a sequence of random bits, *a priori* probability of getting 0 or 1 in any position is $p^f = 1/2$.

Since $S^m$ and $S^f$ are uncorrelated independent sequences $p^m$ and $p^f$ are independent probabilities. The probability of generating a bit 0 or 1, which will construct $k^r$, arising from two independent probabilities $p^m$ and $p^f$ is

$$p^r = p^m . p^f = 1/4 .$$

The probability of generating a bit 0 or 1, which will construct $k^p$ , arising from two independent probabilities $p^m$ and $p^f$ is

$$p^p = p^m . p^f = 1/4.$$

The total probability of generating a bit 0 or 1 which will construct $k^r$ or $k^p$ is

$$p = p^r + p^p = 1/2$$

The generated bits are random bits. Given 4n bits in the input 2n bits are generated in the output. Gain is insignificant. To increase the gain, if we use the generated bits as one of the input and reuse one of the given input, then the generated bits will not be random bits because two inputs (2n+2n) will not be independent inputs. It does not mean that the generated bits cannot be further used to generate bits. Laws of nature tells that that if single generation is possible, then infinite generation is also possible. There is no exception to this natural law. Next we shall present a prototype of the above algorithm which can generate arbitrary numbers of random bits.

## Algorithm- II

Step 0 #

a) Take a sequence $S_0^m$.

b) Note down the random positions of 1s and 0s in the ascending order. It gives two sequences, $K_0^r$, and $K_0^P$ of positive integers. The operation can be symbolically described as, $S_0^m \neg K_0^r$ and $S_0^m \neg K_0^P$.

c) Take another sequence $S_1^f$.

d) Construct complementary sequence $S_2^f$ from $S_1^f$ by interchanging the positions of 0 and 1. Note that $S_1^f$ and $S_2^f$ are maximally correlated random sequences. These two sequences $S_2^f$ and $S_1^f$ have to be used at random.

e) Take another sequence $S^c$ of m random bits. In i-th step, $S_1^f$ or $S_2^f$ will be chosen by the i-th bit of $S^c$.

Step 1 #

a) Choose $S_1^f$ or $S_2^f$ by the 1st bit of $S^C$.

b). Apply $K_0^r$ and $K_0^p$ on the chosen $S_1^f$ or $S_2^f$, and thereby extract a pair of sequences $k_1^r$ and $k_1^p$ of random bits. The operation can be described as $S_0^m \neg K_0^r : (S_1^f / S_2^f) :: k_1^r$ and $S_0^m \neg K_0^p : (S_1^f / S_2^f) :: k_1^p$.

c). Attach $k_1^r$ and $k_1^p$, and thereby construct a new sequence $S_1^m$ of 2n random bits. Here attachment means $k_1^r$ is followed by $k_1^p$. That is, i-th bit of $k_1^r$ is also the i-th bit of the attached string. The attachment can be symbolically described as $S_1^m \equiv k_1^r .. k_1^p$.

d). Note down the positions of 1s and 0s in $S_1^m$. It gives a new pair of sequences $K_1^r$ and $K_1^P$ of positive integers in the ascending order. Note down the positions of 1s and 0s in the string $S_1^m$. The operation can be symbolically described as $S_1^m \neg K_1^r$ and $S_1^m \neg K_1^P$.

Step 2 #

a) Choose $S_1^f$ or $S_2^f$ by the 2nd bit of $S^C$.

b). Apply $K_1^r$ and $K_1^P$ on the chosen $S_1^f$ or $S_2^f$, and thereby extract a new pair of sequences $k_2^r$ and $k_2^p$. The operation can be described as $S_1^m \neg K_1^r : (S_1^f / S_2^f) :: k_2^r$ and $S_1^m \neg K_1^P : (S_1^f / S_2^f) :: k_2^p$.

c). Attach $k_2^r$ and $k_2^p$, and thereby construct a new 2n-bit string $S_2^m$. The attachment can be symbolically described as $S_2^m \equiv k_2^r .. k_2^p$.

d). Note down the positions of 1s and 0s in $S_2^m$. It gives a new pair of sequences $K_2^r$ and $K_2^P$ of positive integers. The operation can be symbolically described as $S_2^m \neg K_2^r$ and $S_2^m \neg K_2^P$.

……………………………………………………………………………………………………….

Step i #

a) Choose $S_1^f$ or $S_2^f$ by the i-th bit of $S^C$.

b). Apply $K_{i-1}^r$ and $K_{i-1}^p$ (taken from previous $S_{i-1}^m$) on the chosen $S_1^f$ or $S_2^f$, and extract a new pair of bit strings $k_i^r$ and $k_i^p$. To get $k_i^r$, bits are extracted from those bit positions in either $S_1^f$ or $S_2^f$ where bits are 1s in $S_{i-1}^m$.To get $k_i^p$, bits are extracted from those bit positions in either $S_1^f$ or $S_2^f$ where bits are 0s in $S_{i-1}^m$.The operations can be symbolically described as $S_{i-1}^m \neg\ K_{i-1}^r : (S_1^f / S_2^f) :: k_i^r$ and $S_{i-1}^m \neg\ K_{i-1}^p : (S_1^f / S_2^f) :: k_i^p$ c). Attach $k_i^r$ and $k_i^p$, and thereby construct a new 2n-bit string $S_i^m$. The attachment can be symbolically described as $S_i^m \equiv k_i^r .. k_i^p$ d). Note down the positions of 1s and 0s in $S_i^m$. It gives a new pair of sequences, $K_i^r$ and $K_i^p$, of positive integers. The operation can be symbolically described as $S_i^m \neg\ K_i^r$ and $S_2^m \neg K_i^p$.

Suppose , in i-th step the integer sequences will be applied on $S_1^f$ if i-th bit of $S^c$ is 0 or on $S_2^f$ if i-th bit of $S^c$ is 1. Let us illustrate the algorithmic steps.

Step 0 #

a). Suppose the following $S_0^m$ is taken.

$$S_0^m = 0\ 1\ 1\ 0\ 1\quad 1\ 0\ 1\ 0\ 0\quad 0\ 1\ 0\ 1\ 1\ 0$$

$$S_0^m = p\ r\ r\ p\ r\quad r\ p\ r\ p\ p\quad p\ r\ p\ r\ r\ p$$

b).

$$K_0^r = \{2, 3, 5, 6, 8, 12, 14,15\}$$

$$K_0^p = \{1, 4, 6,7, 9, 10, 11,13,16\}$$

c). Suppose the following $S_1^f$ is taken.

d).

$$S_1^f = 0\ 1\ 0\ 1\ 1\ 1\ 0\ 1\ 0\ 1\ 0\ 0\ 1\ 0\ 0\ 1$$

$$S_2^f = 1\ 0\ 1\ 0\ 0\ 0\ 1\ 0\ 1\ 0\ 1\ 1\ 0\ 1\ 1\ 0$$

e). Suppose the following $S^c$ is taken. $S^c = \{0 \quad 1 \quad 1 \quad 1 \quad 0 \quad 1 \ldots\ldots\ldots\}$.

Step 1 #

a). Choose $S_1^f$ as 1st bit of $S^c$ is 0.

b). Apply $K_0^r$ and $K_0^p$ on $S_1^f$.

$$\begin{array}{cccccccccccccccccc} S_0^m & = & p & r & r & p & r & r & p & r & p & p & p & r & p & r & r & p \\ S_1^f & = & 0 & 1 & 0 & 1 & 1 & 1 & 0 & 1 & 0 & 1 & 0 & 0 & 1 & 0 & 0 & 1 \end{array}$$

$$\Downarrow$$

$$k_1^r = 1 \quad 0 \quad 1 \quad 1 \quad 1 \quad 0 \quad 0 \quad 0$$

$$k_1^p = 0 \quad 1 \quad 0 \quad 0 \quad 1 \quad 0 \quad 1 \quad 1$$

c). Attach $k_1^r$ and $k_1^p$. The attached sequence $S_1^m \equiv k_1^r \,..\, k_1^p$ is,

$$S_1^m = \{1 \quad 0 \quad 1 \quad 1 \quad 1 \quad 0 \quad 0 \quad 0 \,..\, 0 \, 1 \quad 0 \quad 0 \quad 1 \quad 0 \quad 1 \quad 1\}$$

d). The extracted integer strings are

$$K_1^r = \{1, 3, 4, 5, 10, 13, 15, 16\}$$

$$K_1^p = \{2, 6, 7, 8, 9, 11,12, 14\}$$

Step: 2 #

a). Choose $S_2^f$ as 2nd bit of $S^c$ is 1.

b). Apply $K_1^r$ and $K_1^p$ on $S_2^f$ .

$$\begin{array}{cccccccccccccccccc} S_1^m & \equiv & r & p & r & r & r & p & p & p & p & r & p & p & r & p & r & r \\ S_2^f & \equiv & 1 & 0 & 1 & 0 & 0 & 0 & 1 & 0 & 1 & 0 & 1 & 1 & 0 & 1 & 1 & 0 \end{array}$$

$$\Downarrow$$

$$k_2^r \equiv 1 \quad 1 \quad 0 \quad 0 \quad 0 \quad 0 \quad 1 \quad 0$$
$$k_2^p \equiv 0 \quad 0 \quad 1 \quad 0 \quad 1 \quad 1 \quad 1 \quad 1$$

c). Attach $k_2^r$ and $k_2^p$. The attached sequence $S_2^m \equiv k_2^r \,..\, k_2^p$ is,

$$S_2^m = \{1 \quad 1 \quad 0 \quad 0 \quad 0 \quad 0 \quad 1 \quad 0 \,..\, 0 \, 0 \quad 1 \quad 0 \quad 1 \quad 1 \quad 1 \quad 1\}$$

d). The extracted integer strings are

$$K_2^r = \{1, 2, 7, 11, 13, 14, 15, 16\}$$

$$K_2^p = \{3, 4, 5, 6, 8, 9, 10, 12\}$$

Let us now prove that in each step the algorithm-II generates random bits.

***Proof :*** Since $S_1^f$ is a sequence of random bits *a priori* probability of getting 0 or 1 in j-th position is $p_1^f$ =1/2. Since $S_2^f$ is complimentary to $S_1^f$ , *a priori* probability of getting 0 or 1 in the same j-th position is $p_2^f$ =1/2. Note that $S_1^f$ and $S^c$ are uncorrelated independent sequences. Similarly, $S_2^f$ and $S^c$ are uncorrelated independent sequences. In each step, $S_1^f$ or $S_2^f$ is chosen by a bit of $S^c$ with probability $p^c$=1/2. Therefore, total probability of getting 0 or 1 in any position is,

$$p^f = p^c . p_1^f + p^c . p_2^f = 1/2$$

Since $S_0^m$ is a sequence of random bits, *a priori* probability of getting 0 or 1 is $p_0^m$= 1/2. Since $S_0^m$ and $S^c$ are uncorrelated independent sequences $p_0^m$ and $p^f$ are independent probabilities. The probability of generating a bit 0 or 1, which will construct $k_1^r$, arising from two independent probabilities $p_0^m$ and $p^f$ is

$$p_1^r = p_0^m . p^f = 1/4$$

The probability of generating a bit 0 or 1, which will construct $k_1^p$, arising from two independent probabilities $p_0^m$ and $p^f$ is

$$p_1^p = p_0^m . p^f = 1/4$$

The total probability of generating a bit 0 or 1 which will construct $k_1^r$ or $k_1^p$ is

$$p_1 = p_1^r + p_1^p = 1/2$$

Since extracted pair constructs $S_1^m$, $p_1$ can be denoted as $p_1^m$. Since $S_1^m$ and $S^c$ are uncorrelated independent sequences $p_1^m$ and $p^f$ are independent probabilities. The probability of generating a bit 0 or 1, which will construct $k_2^r$, arising from two independent probabilities $p_1^m$ and $p^f$ is

$$p_2^r = p_1^m . p^f = 1/4 .$$

The probability of generating a bit 0 or 1, which will construct $k_2^p$, arising from two independent probabilities $p_1^m$ and $p^f$ is

$$p_2^p = p_1^m . p^f = 1/4 .$$

The total probability of generating a bit 0 or 1 which will construct $k_2^r$ or $k_2^p$ is,

$$p_2 = p_2^r + p_2^p = 1/2$$

Since extracted pair of (i-1)-th step constructs $S_{i-1}^m$, $p_{i-1}$ can be denoted as $p_{i-1}^m$. Since $S_{i-1}^m$ and $S^c$ are uncorrelated independent sequences $p_{i-1}^m$ and $p^f$ are independent probabilities. The probability of generating a bit 0 or 1, which will construct $k_i^r$, involving two independent probabilities $p_{i-1}^m$ and $p^f$ is

$$p_i^r = p_{i-1}^m . p^f = 1/4$$

The probability of generating a bit 0 or 1, which will construct $k_i^p$, involving two independent probabilities $p_{i-1}^m$ and $p^f$ is

$$p_i^p = p_{i-1}^m . p^f = 1/4$$

The total probability of generating a bit 0 or 1 which will construct $k_i^r$ and $k_i^p$ is,

$$p_i = p_i^r + p_i^p = 1/2$$

In each step algorithm generates random bits.

Some of the features of the algorithm are discussed below.

1. An attached string can become identical with $S_1^f$ or $S_2^f$ or $S_0^m$. If any attached string becomes identical with $S_1^f$ or $S_2^f$ the algorithm will halt. If attached string becomes identical with $S_0^m$ the algorithm will be cyclic. The probability of generating identical $S_1^f$ or $S_2^f$ or $S_0^m$ is $1/2^{2n}$.

2. In R steps the total number of generated random bits is 2nR while the number of input random bits is 4n + R. Gain is, G = 2nN – (4n +R). Some of the generated random bits can be used to increase the length of the input strings $S_1^f$ $(S_2^f)$, $S_0^m$ and $S^c$. The algorithm can ultimately run on two arbitrarily long strings.

3. As far as correlation is concerned there are mainly two possibilities.

a) In each step, independent inputs are given and there is no correlation between the inputs of i-th steps and j-th steps.

In this case, i-th input can be correlated with the i-th output. This is not n-bit n-bit correlation. This is n-bit 2n-bit correlation. But i-th output cannot be correlated with j-th output and i-th input cannot be correlated with j-th output. So n-bit outputs can remain uncorrelated even they are correlated with 2n-bit inputs.

b) In each step, independent inputs are given but there is a correlation between the inputs of i-th steps and j-th steps. The algorithm II is an example.

In this case, i-th input can be correlated with the i-th output and i-th input cannot be correlated with the j-th output, but i-output cannot be correlated with j-th output. But this correlation can be known if and only if the input strings are known. Therefore, so far output is concerned this is a *meaningless* correlation. Due to this *meaningless* correlation the same output can be generated by giving the same input. This is advantageous for Monte-Carlo simulation when simulation needs to be repeated under the same condition. One may tell it as *meaningfully meaningless* correlation.

4. In the output the probable number of different 2n-bit sequences is $2^{2n}$. For all probable input pairs the probable number of different order of 2n-bit sequences of sequence is $\left(2^{2n}\right)^{2^{2n}}$. For a pair of particular inputs $S_1^f$ and $S_0^m$ the probable number of different order of 2n-bit sequences of sequence is $2^{2^{2n}}$. A pair of attached strings can be used as a new pair of inputs. Therefore, starting with initial pair all probable $\left(2^{2n}\right)^{2^{2n}}$ sequences of sequences can be generated.

The probability of generating any 2n-bit sequence is $1/2^{2n}$. The probability of generating the same 2n-bit sequence is also $1/2^{2n}$. The probability of generating any sequence of 2n-bit sequences is $\frac{1}{\left(2^{2n}\right)^{2^{2n}}}$. The same output sequences can be generated if the same inputs are given. This is advantageous. Random events cannot generate the same distribution if we demand so.

5. Given the pair of output sequence it is impossible to identify the pair of input sequences. Because the same pair of output sequences can be generated by ${}_2C^{2n}$ pairs of input sequences. But the question is, given the sequence of outputs of the algorithm is it possible to identify the algorithm ? This is not possible if we can find another prototype of this algorithm. One of the prototype of the presented algorithm is discussed below.

Instead of one sequence two sequences $S_0^m$ and $S_0^m$ and can be simultaneously applied either on $S_1^f$ or $S_2^f$ at random to generate four sequences of random bits taking the bits from those positions of $S_1^f$ or $S_2^f$ where bits are : 1). 1s in $S_0^m$ and 0s in $S_0^m$ 2). 0s in $S_0^m$ and 1s in $S_0^m$ 3). 1s in $S_0^m$ and 1s in $S_0^m$ 4). 0s in $S_0^m$ and 0s in $S_0^m$ . Total probability of generating a bit will be 1/8 +1/8 +1/8 + 1/8 =1/2. As if three coins jointly generate a bit with probability 1/2.

We conjecture that exponentially large numbers of prototypes of this algorithm do exist and can be found by exponentially large numbers of researchers.

6. The algorithm can also be used to generate non-random numbers. If one of the three input sequences contains non-random numbers the algorithm will generate non-random numbers.

The algorithm generates sequence of two equally probable numbers $X_i$ where i = 1,2. Following the same algorithmic technique sequence of N equally probable numbers $X_i$ can be generated where i =1,2,….N. In each step, N sequences, each consists of N equally probable numbers $X_i$, can be generated if previously generated N sequences are attached and then the attached sequence $S_i^m$ of length Nn is applied on any one of the N! complimentary sequences $S_1^f, S_2^f, S_3^f$,....... , each of length Nn, at random. It appears that in each step, to select N! sequences (at random) with equal probability 1/N! we need a sequence $S^c$ of N! equally probable numbers. But the problem is, we need an arbitrarily long sequence $S^c$ of N! equally probable numbers to generate arbitrarily long sequence of N equally probable numbers. This is not an insurmountable problem. We have a found a solution which is discussed below.

The requirement of arbitrarily long sequence $S^c$ of N! equally probable numbers can be avoided if we first generate sequence $S^{c_2}$ of 2 equally probable numbers followed by sequence $S^{C_3}$ of 3 equally probable numbers and finally sequence $S^{C_{N-1}}$ of (N-1) equally probable numbers by this algorithmic technique. Along with this sequences another sequence $S^{C_N}$ of N equally probable numbers can be taken. Now if we take the i-th numbers of the sequences $S^{C_2}, S^{C_3} \ldots . S^{C_{N-1}}, S^{C_N}$, then N! distributions are possible. Each arrangement $(X_2)_i\ (X_3)_i\ (X_4)_i \ldots (X_N)_i$ can be used to choose any one of the sequences $S_1^f, S_2^f, S_3^f, \ldots\ldots. S_{N!}^f$ with equal probability 1/N!. Of course which sequence $S_i^f$ will be chosen by which probable distribution that instruction should be given in the input. To generate a sequence of N probable numbers we have to generate sequences of first 2, then 3… and finally N-1 equally probable numbers. Generating 2, 3, 4.. N equally probable numbers they can be used to increase the length of the sequences $S^{C_2}$, $S^{C_3}, S^{C_3} \ldots . S^{C_N}$ respectively

<u>Generalized Algorithm</u>

Step 0 #

a). Take a sequence $S_0^m$ of N equally probable numbers $X_i$ where the length of $S_0^m$ is Nn.

b). Note down the random positions of $X_1$, $X_2$, $X_3$ … $X_i$ in the ascending order which gives sequences, $K_0^1, K_0^2 \ldots .. K_0^i$ of positive integers. The operation can be described as

$S_0^m \neg K_0^1, S_0^m \neg K_0^2, \ldots\ldots\ldots S_0^m \neg K_0^i$.

c). Take another sequence $S_1^f$ of N equally probable numbers $X_i$ where the length of $S_1^f$ is Nn.

d). From $S_1^f$, construct complimentary sequences $S_2^f, S_3^f, \ldots\ldots.$ of $S_1^f$ by interchanging the positions of $X_i$. Since N numbers can be arranged in N! ways so positions of N numbers can be interchanged in N!

ways. Therefore, N! complimentary sequences $S_i^f$ of N equally probable numbers are possible. If one of these complimentary sequences is known then the others can be constructed.

e). Take a set of sequences $S^c \in (S^{C_2}, S^{C_3}, S^{C_3} \ldots S^{C_N})$ of 2, 3, … (N-1), N probable numbers.

Step i #

a) Choose $S_1^f$ or $S_2^f$ or $S_3^f$……or $S_i^f$ by the arrangement $(X_2)_i$ $(X_3)_i$ $(X_4)_i$ …$(X_N)_i$ of i-th number of each of the sequences $S^{C_2}, S^{C_3}, S^{C_3} \ldots, S^{C_N}$.

b). Apply $K_{i-1}^1$, $K_{i-1}^2$……$K_{i-1}^i$ (taken from previous $S_{i-1}^m$) on the chosen $S_1^f$ or $S_2^f$ or $S_3^f$……or $S_i^f$ and extract the sequences $k_i^1$, $k_i^2$……$k_i^i$. The operation can be symbolically described as

$$S_{i-1}^m \neg\ K_{i-1}^1 : (S_1^f / S_2^f / S_3^f \ldots / S_i^f) :: k_i^1$$

$$S_{i-1}^m \neg\ K_{i-1}^2 : (S_1^f / S_2^f / S_3^f \ldots / S_i^f) :: k_i^2$$

$$\ldots\ldots\ldots\ldots\ldots\ldots\ldots\ldots\ldots\ldots\ldots\ldots\ldots$$

$$S_{i-1}^m \neg\ K_{i-1}^i : (S_1^f / S_2^f / S_3^f \ldots / S_i^f) :: k_i^i.$$

In this operation two numbers XX are used to generate a single number X where the first number X belongs to $S_{i-1}^m$ and the second number X belongs to $S_i^f$.

The coding can be described as

$$\text{For } k_i^1,\quad X_1X_1 \equiv X_1,\ X_1X_2 \equiv X_2,\ \ldots\ldots,\ X_1X_i \equiv X_i$$

$$\text{For } k_i^2,\quad X_2X_1 \equiv X_1,\ X_2X_2 \equiv X_2,\ \ldots\ldots,\ X_2X_i \equiv X_i$$

$$\ldots\ldots\ldots\ldots\ldots\ldots\ldots\ldots\ldots\ldots\ldots\ldots\ldots\ldots\ldots\ldots$$

$$\text{For } k_i^2,\quad X_iX_1 \equiv X_1,\ X_iX_2 \equiv X_2,\ \ldots\ldots,\ X_1X_i \equiv X_i$$

c). Attach $k_i^1$, $k_i^2$……$k_i^i$, and thereby construct a new sequence $S_i^m$. The operation can be symbolically described as $S_i^m \equiv k_i^1 .. k_i^2 .. k_i^3 \ldots\ldots k_{i-1}^i .. k_i^i$ d). Note down the positions of $X_1$, $X_2$, $X_3$ … $X_i$ in $S_i^m$. It gives a new sequences, $K_i^1$, $K_i^2$…… $K_i^i$ of positive integers. The operation can be symbolically described as $S_i^m \neg K_i^1, S_i^m \neg K_i^2, \ldots\ldots\ldots S_i^m \neg K_i^i$

As for example, for the generation of arbitrarily long string of 3 equally probable numbers we need inputs: $S_0^m$, $\{S_1^f, S_2^f, S_3^f, S_4^f, S_5^f, S_6^f\}$, $\{S_2^C, S_3^C\}$. Therefore, we have to first generate arbitrarily long sequence $S_2^C$ of 2 equally probable numbers to generate arbitrarily long sequence $S_3^C$ of 3 equally probable numbers by this algorithm. Here, out of 3! arrangements one arrangement $(X_2)_i$ $(X_3)_i$ of i-th numbers of $S_2^C$ and $S_3^C$ will determine one of the six sequences $S_1^f, S_2^f, S_3^f, S_4^f, S_5^f, S_6^f$ on which $S_0^m$ will be applied with probability 1/6 . Note that 2 equally probable numbers are the primary seeds.

The probability of generating any sequence $S_i^m$ of N equally probable numbers $X_i$ is $\frac{1}{N^{Nn}}$. The probability of generating the same sequence $S_i^m$ is also $\frac{1}{N^{Nn}}$. The probability of generating any sequence of sequences $S_i^m$ is $\frac{1}{\left(N^{Nn}\right)^{N^{Nn}}}$. Generated strings cannot be correlated among themselves. Given a string of random number it is not possible to tell how it was generated.

Like previous case, it is easy to see that this generalized algorithm generates a number with probability 1/N.

***Proof***: The probability of selecting any one of the N! sequences $S_1^f, S_2^f, S_3^f, \ldots\ldots. S_{N!}^f$ by the arrangement of i-th numbers of the sequences $S^{C_2}$, $S^{C_3}, S^{C_3} \ldots, S^{C_N}$ is $p^c = \frac{1}{2}.\frac{1}{3}.\frac{1}{4}\ldots\ldots\frac{1}{N} = \frac{1}{N!}$. Since $S_1^f$ is sequence of random numbers *a priori* probability of getting a number $X_1$ in j-th position is $P^1$ = 1/N. Since $S_2^f$ is a complimentary sequence of $S_1^f$ *a priori* probability of getting $X_2$ in the same j-th position of the sequence $S^f_2$ is $P^2$ = 1/N. Since $S^f_{N!}$ is a complimentary sequence of $S^f_1$ *a priori* probability of getting $X_N$ in the same j-th position of the sequence $S^f_{N!}$ is $P^{N!}$ = 1/N . Note that the set of sequences $S^c$ $\in (S^{C_2}, S^{C_3}, S^{C_3} \ldots, S^{C_N})$ is independent from the set of sequences $S^f \in (S^f_1, S^f_2, S^f_3, \ldots\ldots. S^f_{N!}\ S^f_i)$. In each step, one sequence $S^f_i$ is chosen by the random number of $S^c$ with probability $p^c$ =1/N. Total probability of getting a number $X_i$ in any position is,

$$p^f = \sum_1^{N!} (P^1.p^c + P^2.p^c + \ldots.. + P^{N!}.p^c)$$

$$= \sum_1^{N!} (1/N!.1/N + 1/N!.1/N + 1/N!\ .1/N\ldots..1/N!.1/N)$$

$$= 1/N$$

Since $S^m_{i-1}$ is a sequence of random numbers *a priori* probability of getting each number $X_i$ in $S^m_i$ is $p^m_{i-1}$ = 1/N. Since $S^m_{i-1}$ and $S^c$ are uncorrelated independent sequences $p^m_{i-1}$ and $p^f$ are independent probabilities. The probability of generating a number, which will construct $k^i_i$, arising from two independent probabilities $p^m_{i-1}$ and $p^f$ is

$$p^N_i = p^m_{i-1}.p^f = 1/N^2$$

Total probability of generating a number $X_i$ which will construct $k^1_i$ or $k^2_i$ or …or $k^i_i$ is,

$$p_i = p^1_i + p^2_i + \ldots\ldots. p^N_i$$

$$= \sum_1^N (1/N^2 + 1/N^2 + \ldots..+1/N^2)$$

$$= 1/N$$

Random numbers are extensively used in Monte Carlo simulation and for cryptographic purposes [2]. The algorithm can be used for these purposes. The algorithm can be used to generate unbreakable message, PIN and password. Even common people can use the algorithm for these purposes. The cryptographic use of the algorithm has been discussed [4,5] elsewhere.

The algorithms reveals that there is a numerical origin of randomness. Needless to say, the algorithms can be implemented on classical computer. The existence of these algorithms implies that classical computer and human brain can behave like a probabilistic machine. It may be recalled that Feynman envisioned [6] quantum computer for the simulation of quantum system. Although classical computer is not suitable for the simulation of quantum system, but intrinsic randomness of quantum system can be simulated on classical computer.

The existence of probabilistic algorithms [7,8] including Monte-Carlo based algorithms [2] gave hint that classical computer cannot be considered as a totally deterministic machine. The presented algorithm corroborates that observation.

Note added: Interested readers may see some wekly related works [9-12].

email: mitra1in@yahoo.com

.